\documentclass[%
aip,
apl,
superscriptaddress,
 amsmath,amssymb,
 reprint,%
]{revtex4-1}

\usepackage{graphicx}
\graphicspath{{Figures/},{Figures/Appendix/}}
\usepackage{dcolumn}
\usepackage{bm}

\usepackage[utf8]{inputenc}
\usepackage[T1]{fontenc}
\usepackage{mathptmx}
\usepackage{upgreek}
\usepackage{lipsum}
\usepackage{todonotes}
\usepackage{subcaption}
\usepackage{float}
\usepackage{multirow}
\usepackage{adjustbox}
\usepackage{cleveref}
\usepackage{IEEEtrantools}
\usepackage{braket}

\begin{document}
\preprint{AIP/123-QED}

\title{Low-frequency spin qubit detuning noise in highly purified $^{28}$Si/SiGe}

\author{Tom Struck}%
\affiliation{JARA-FIT Institute for Quantum Information, Forschungszentrum J\"ulich GmbH and RWTH Aachen University, Aachen, Germany}
\author{Arne Hollmann}%
\affiliation{JARA-FIT Institute for Quantum Information, Forschungszentrum J\"ulich GmbH and RWTH Aachen University, Aachen, Germany}
\author{Floyd Schauer}%
\affiliation{Institut für Experimentelle und Angewandte Physik, Universit\"at Regensburg, Regensburg, Germany}
\author{Olexiy Fedorets}%
\affiliation{JARA-FIT Institute for Quantum Information, Forschungszentrum J\"ulich GmbH and RWTH Aachen University, Aachen, Germany}
\author{Andreas Schmidbauer}%
\affiliation{Institut für Experimentelle und Angewandte Physik, Universit\"at Regensburg, Regensburg, Germany}
\author{Kentarou Sawano}%
\affiliation{Advanced Research Laboratories, Tokyo City University, Tokyo, Japan}
\author{Helge Riemann}%
\affiliation{Leibniz-Institut für Kristallzüchtung (IKZ), Berlin, Germany}
\author{Nikolay V. Abrosimov}%
\affiliation{Leibniz-Institut für Kristallzüchtung (IKZ), Berlin, Germany}
\author{{\L}ukasz Cywi{\'n}ski}%
\affiliation{Institute of Physics, Polish Academy of Sciences, Warsaw, Poland}
\author{Dominique Bougeard}%
\affiliation{Institut für Experimentelle und Angewandte Physik, Universit\"at Regensburg, Regensburg, Germany}
\author{Lars R. Schreiber}%
\affiliation{JARA-FIT Institute for Quantum Information, Forschungszentrum J\"ulich GmbH and RWTH Aachen University, Aachen, Germany}
\email[email to: ]{lars.schreiber@physik.rwth-aachen.de}

\begin{abstract}
The manipulation fidelity of a single electron qubit gate-confined in a $^{28}$Si/SiGe quantum dot has recently been drastically improved by nuclear isotope purification. Here, we identify the dominant source for low-frequency qubit detuning noise in a device with an embedded nanomagnet, a remaining $^{29}$Si concentration of only 60\,ppm in the strained $^{28}$Si quantum well layer and a spin echo decay time $T_2^{\text{echo}}=128\,\upmu$s. The power spectral density (PSD) of the charge noise explains both the observed transition of a $1/f^2$- to a $1/f$-dependence of the detuning noise PSD as well as the observation of a decreasing time-ensemble spin dephasing time from $T_2^* \approx 20\,\upmu$s with increasing measurement time over several hours. Despite their strong hyperfine contact interaction, the few $^{73}$Ge nuclei overlapping with the quantum dot in the barrier do not limit $T_2^*$, as their dynamics is frozen on a few hours measurement scale. We conclude that charge noise and the design of the gradient magnetic field is the key to further improve the qubit fidelity.
\end{abstract}

\maketitle
Gate-defined quantum dots (QDs) are a promising platform to confine and control single spins, which can be exploited as quantum bits (qubits) \cite{RevModPhys.85.961}. Unlike charge, a single spin does not couple directly to electric noise. Dephasing is dominated by magnetic noise, typically from the nuclear spin bath overlapping with the QD \cite{assali_hyperfine_2011}. The use of silicon as a qubit host material boosted the control of individual spins by minimizing this magnetic noise: in addition to the intrinsically low hyperfine interaction in natural silicon, the existence of nuclear spin-free silicon isotopes, e.g. $^{28}$Si, allows isotopical enrichment in crystals \cite{Abrosimov_2017,Itoh_2014}. Controlling individual electrons and spins in highly enriched $^{28}$Si quantum structures \cite{Wild2012, Muhonen2014, Veldhorst2014, Lawrie_2019} then opens the door to an attractive spin qubit platform realized in a crystalline nuclear spin vacuum. Indeed, two-qubit gates \cite{Veldhorst2015,Zajac439,Watson2018} have recently been demonstrated in natural and enriched quantum films, while isotopical purification of $^{28}$Si down to 800 ppm of residual nuclear spin-carrying $^{29}$Si allowed to push manipulation fidelities beyond 99.9\% for a single qubit  \cite{tarucha_quantum-dot_2017, Yang19} and towards 98\% for two qubits \cite{Huang2019}.
Qubit manipulation of individual spins is currently either realized with local AC magnetic fields generated by a stripline to drive Rabi transitions \cite{Koppens2006, Muhonen2014, Veldhorst2014} or via artificial spin-orbit coupling engineered by a micromagnet integrated into the device. This latter approach is advantageous by allowing the control of spin qubits solely by local AC electric fields \cite{Pioro-Ladriere2008,tarucha_quantum-dot_2017,Watson2018,Zajac439}, permitting excellent local control and faster Rabi frequencies. At the same time it opens a new dephasing channel for electric noise, due to the static longitudinal gradient magnetic field of the micromagnet, competing with the magnetic noise. To fully exploit the potential of magnetic noise minimization through isotope enrichment in $^{28}$Si/SiGe, two experimental questions thus become relevant for devices with integrated static magnetic field gradients: Firstly, to what extent electronic noise impacts the spin qubit dephasing compared to magnetic noise \cite{Zhao_2019} and, secondly, which role the natural SiGe potential wall barriers play for dephasing, since the hyperfine interaction of bulk Ge exceeds the one of bulk Si by a factor of approximatively 100 \cite{PhysRev.134.A265, PhysRevB.85.205312}.

Here, we present an electron spin qubit implemented in a highly isotopically purified $^{28}$Si/SiGe device, with only 60\,ppm of residual $^{29}$Si, which includes a magnetic field gradient generated by a nanomagnet integrated into the electron-confining device plane. We use Ramsey fringe experiments to investigate the detuning noise spectrum of the single electron spin down to 10$^{-5}$~Hz. We find the frequency dependence of the qubit detuning spectrum to be identical to the spectrum of the device's electric charge noise over more than 8 decades. At low frequencies, below $5\cdot10^{-3}$\,Hz, both noise spectra decrease with $1/f^2$. Above, they transit to a $1/f$ dependence and finally present a behavior comparable to a device\cite{tarucha_quantum-dot_2017} featuring a micromagnet and 800\,ppm $^{29}$Si  at higher frequencies, as deduced from a Hahn-echo sequence for the detuning, yielding $T_2^\text{echo}=128\,\upmu \text{s}$. Electric noise thus dominates our qubit dephasing in a broad frequency range. It is also responsible for the observed decrease of $T_2^*$ with increasing measurement time\cite{Dial13}. Interestingly, although we show the $^{73}$Ge in the quantum well-defining natural SiGe to represent a potential limitation for our device, our experiments suggest the nuclear spin bath to be frozen on a time scale of hours and to much less contribute to $T_2^*$ than expected at the ergodic limit.

\begin{figure}[ht!]
    \includegraphics[width=\linewidth]{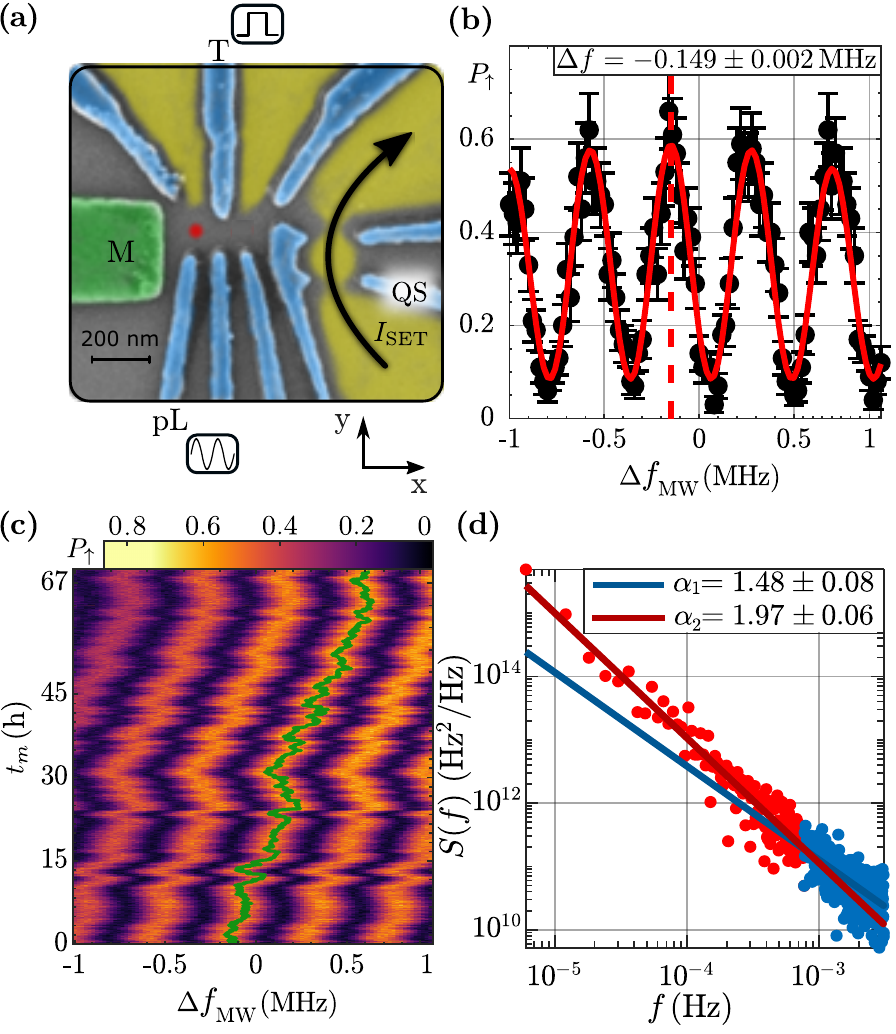}   
    \caption{ (a) Colored scanning electron micrograph of a sample similar to the one used in this work.(b) Measurement of Ramsey fringes. The spin-up probability is recorded as a function of the resonance detuning $\Delta f_{MW}$. Each point corresponds to 100 single-shot measurements. The position of the spin resonance is indicated by the dashed red line.
    (c) Time evolution of the Ramsey fringe pattern during a measurement time $t_m=67$~hours. The green solid line tracks the resonance detuning $\Delta f$ extracted from the fringes.
    (d) PSD $S(f)$ of the qubit detuning calculated from the data shown in panel (c).}
    \label{fig:measurement}
\end{figure}

The device used for all measurements consists of an undoped $^{28}$Si/SiGe heterostructure confining a two dimensional electron gas in $^{28}$Si with 60\,ppm of residual $^{29}$Si. Metal gates are used to form a quantum dot (QD) containing a single electron (Fig. \ref{fig:measurement}(a)) . The charge state of the QD is detected via a single electron transistor (SET) located at the right-hand side of the device. The large gate labeled M on the left-hand side is a single domain Co nanomagnet. Its stray-magnetic field provides a magnetic field gradient \cite{Petersen2013} for spin driving by electric dipole spin resonance (EDSR). For details of the device see the supplements and Ref. \cite{Hollmann_2019}. We apply an external magnetic field of 668~mT along the $x$-direction.

First, we focus on the PSD of the frequency detuning $\Delta f$ of the qubit with respect to a reference frequency of $f_R=19.9\,$GHz. $\Delta f$ is determined by a Ramsey fringe measurement, during which the microwave pulses are detuned from the reference by $\Delta f_{MW}$ (Fig. \ref{fig:measurement}(b)). We vary $\Delta f_{MW}$ from $-1$ to $1$\,MHz in 100 steps. Each point of the spin-up probability $P_{\uparrow}$ is an average over 100 single-shot measurements. One Ramsey fringe, which is one measurement of $\Delta f$, takes 120~s. We fit $\Delta f$ by applying the formula for the fringe pattern \cite{PhysRevB.83.235201}:

\begin{IEEEeqnarray}{rCl}\label{eq:RF}
	&P&_{\uparrow}(f_{R},t_e,\Delta f,t_{\frac{\pi}{2}})  = \frac{4 f_{R}^2}{\Phi^2} \cdot \sin \left(\pi t_{\frac{\pi}{2}} \Phi \right)^2 \cdot
	\\
	&\Bigg[& \cos(\pi \Delta f t_e) \cdot \cos \left( \pi t_{\frac{\pi}{2}} \Phi \right) 
	- \frac{\Delta f}{\Phi} \cdot \sin(\pi \Delta f t_e)
	\cdot \sin \left( \pi t_{\frac{\pi}{2}} \Phi \right) \Bigg]^2 \nonumber
\end{IEEEeqnarray}
where $\Phi= \sqrt{\Delta f^2 + f_{R}^2}$, $t_e$ is the evolution time between the two $\pi/2$ gates and $t_{\frac{\pi}{2}}$ is the execution time of the $\pi/2$ gate. 

Fig. \ref{fig:measurement}(c) displays Ramsey fringes recorded during a measurement time period $t_m$ of 67 hours. The green line tracks $\Delta f$ during the full time period. We calculated the PSD $S(f)$ of the qubit detuning with Welch's method (Fig. \ref{fig:measurement}(d)). For frequencies below $\approx 7\cdot10^{-4}$~Hz, we find a $S(f)\propto1/f^{1.97}$ dependence. It transitions into a region with smaller exponent, here fitted with $S(f)\propto1/f^{1.48}$ (blue line in Fig. \ref{fig:measurement}(d)). Note that spin qubits in GaAs which dephase dominantly due to hyperfine interaction\cite{Hanson07,Cywinski09,Cywinski11} are also characterized by a $1/f^2$ dependence in their low-frequency detuning noise PSD, which has been assigned to nuclear spin diffusion there.\cite{Reilly08, Malinowski17}

\begin{figure}
    \includegraphics[width=\linewidth]{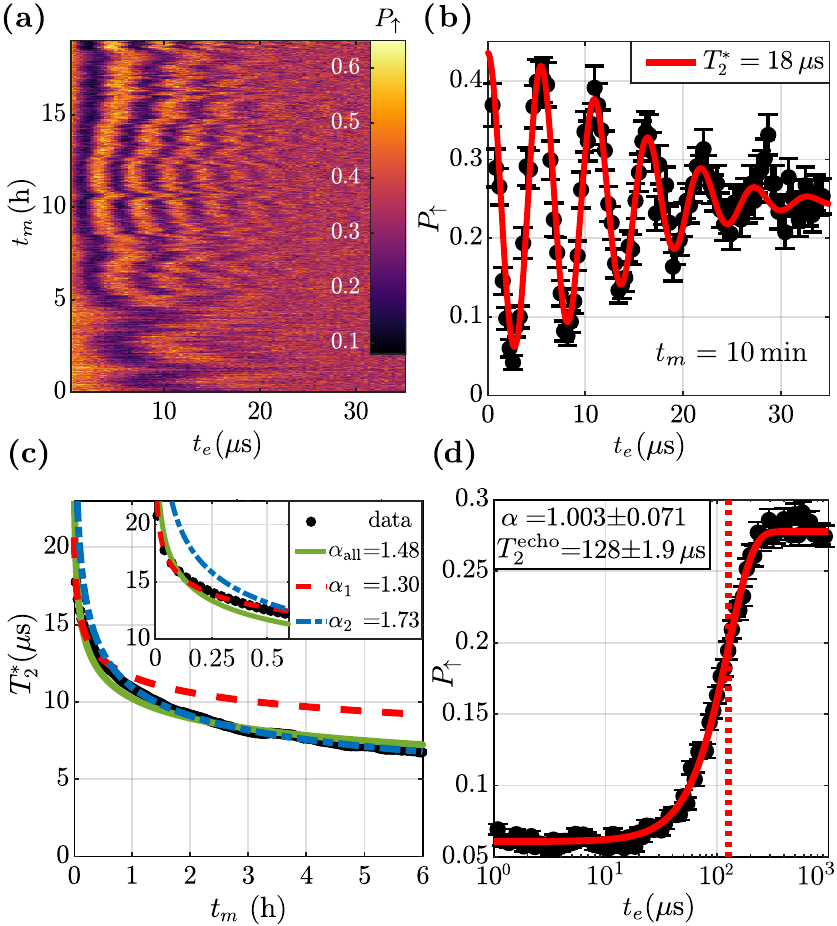}
    \caption{
    (a) Time evolution of the $T_2^*$ measurement. Each point is an average over 50 single-shot measurements.
    (b) Measurement of the spin-up probability as a function of the evolution time $t_e$. Each point corresponds to 500 single-shot measurements. The solid line shows a fit of a Gaussian decay revealing the time-ensemble dephasing time $T_2^*$.
    (c) Dependence of $T_2^*$ on the measurement time. The solid green line shows a fit to all data points with one $\alpha$-value. The red and blue lines show fits to the long and shorter measurement times.
    (d) Spin-up probability as a function of the evolution time $t_e$ after a Hahn-echo gate sequence. Each point is an average over 5000 single shot measurements. The solid line is a fit to Eq. \ref{eq:echo_fit}. The dashed line marks the fitted $T_2^{\text{echo}}$.}
    \label{fig:fig2}
\end{figure}

Having analyzed the qubit detuning noise $S(f)$ in the low-frequency regime, we now investigate its impact on the time-ensemble spin dephasing time $T_2^*$. We recorded $P_{\uparrow}$ during a series of Ramsey sequences with varying $t_e$ (Fig. \ref{fig:fig2}(a)) in every line. For each $t_m$, we average as many consecutive $P_{\uparrow}(t_e)$ lines of this dataset as required to reach a total measurement time $t_m$. The averaged $P_{\uparrow}(t_e)$ was then fitted with 
\begin{equation}
    P_{\uparrow}(t_e)=A\cdot \exp\left(-\left(\frac{t_e}{T_2^*}\right)^2\right)\cos(2 \pi \Delta f \cdot t_e)+B,
\end{equation}
\noindent where $A$ and $B$ are constants related to the qubit initialization and readout fidelity. An example of $P_{\uparrow}(t_e)$ measured over $t_m=10$~min is shown in Fig. \ref{fig:fig2}(b). We extract $T_2^*(t_m=10~$min$)=18\, \upmu$s. To achieve better statistics, this procedure was executed consecutively for different bundles of $P_{\uparrow}(t_e)$ lines. We chose to offset the bundles by 25 lines giving overlap between them.
This results in each $T_2^*(t_m)$ value being averaged from $~900$ $T_2^*(t_m)$ values using different line bundles from the dataset displayed in Fig. \ref{fig:fig2}(a) and a second dataset not shown here. Fig. \ref{fig:fig2}(c) shows these averaged $T_2^*(t_m)$ for $t_m$ ranging between 38~seconds and 6.3~hours. 
Remarkably, $T_2^*(t_m)$ drops monotonously with increasing measurement time without saturating for long $t_m$, qualitatively matching the qubit detuning noise PSD $S(f)$, which keeps increasing towards low frequencies (Fig. \ref{fig:measurement}(d)). In a rough approximation, considering detuning noise of the type $S(f)=S_0/f^{\alpha}$ with $\alpha \gtrsim 1$, $T_2^*(t_m)$ induced by detuning noise is (see Supplement Notes 1):

\begin{equation}
    T_2^*(t_m)=\left(\frac{4\pi^2S_0}{\alpha-1}\left( t_m^{\alpha-1}-t_e^{\alpha-1} \right)\right)^{-\frac{1}{2}}.
    \label{eq:t2star_formula}
\end{equation}

 \noindent Fitting $T_2^*(t_m)$ with only one $\alpha_{\text{all}}=1.48$ (green solid line) shows clear deviation from the data points (Fig. \ref{fig:fig2}(c)). Motivated by the variation of $\alpha$ in $S(f)$, we fit two separate ranges of $t_m$ above and below $t_m=25\,\text{min}$ (that is $6.7\cdot10^{-4}$~Hz, which is very close to the transition point $7\cdot10^{-4}$~Hz found in Fig. \ref{fig:measurement}(d)), which are characterized by $\alpha_1=1.30$ (red dashed curve) and $\alpha_2=1.73$ (blue dashed curve), and are in good qualitative agreement with $\alpha_1$ and $\alpha_2$ found for the detuning noise in Fig. \ref{fig:measurement}(d). The quantitative deviation of both $\alpha_i$ ($i=1,2$) determined by the $T_2^*(t_m)$ compared to the ones directly fit to the PSD results from the fact that the $T_2^*$ measurement integrates over the PSD from $t_m^{-1}$ to $t_e^{-1}$.
 
Our spin-detection bandwidth in the Ramsey fringe experiment sets a limit on the maximum frequency of $S(f)$ in Fig. \ref{fig:measurement}(d). To gain information on $S(f)$ at a higher frequency, we performed a Hahn-echo experiment, that is extended the Ramsey control sequence by a $\pi_X$ gate between the two $(\pi/2)_X$ gates, in order to filter out low frequency noise. The measured data (Fig. \ref{fig:fig2}(d)) has been fitted with   

\begin{equation}
    P_{\uparrow}(t_e)=A\cdot\left(1-\exp{\left(-\left(\frac{t_e}{T_2^{\text{echo}}}\right)^{\alpha+1}\right)} \right)+B.
    \label{eq:echo_fit}
\end{equation}

\noindent We find $\alpha=1.003\pm 0.071$ and $T_2^{\text{echo}}=128\pm1.9\,\upmu$s. We can deduce that $S(f) \propto 1/f$ at a frequency of approximately $f=1/T_2^{\text{echo}}=7.8$~kHz, in line with the observations in a device with an on-chip micromagnet and 800\,ppm residual $^{29}$Si for $f>10^{-2}$~Hz \cite{tarucha_quantum-dot_2017}. With the low-frequency PSD (Fig. \ref{fig:measurement}(d)) and these spin echo results we conclude that the initial $S(f) \propto 1/f^2$ dependence observed at low frequencies transits to a $S(f) \propto 1/f$ dependence around $7\cdot10^{-4}$~Hz to $1\cdot10^{-3}$~Hz. With the detection bandwidth limit set by the Ramsey fringe experiment at approximately $3 \cdot 10^{-3}$~Hz, we observe this gradual transition, explaining $\alpha=1.48$ found in Fig. \ref{fig:measurement}(d). Remarkably, we find a 28~\% higher $T_2^{\text{echo}}$ compared to the device with an on-chip micromagnet and 800\,ppm residual $^{29}$Si \cite{tarucha_quantum-dot_2017}, indicating that overall the detuning noise is lower in our sample in this regime. 

\begin{figure}
    \includegraphics[width=\linewidth]{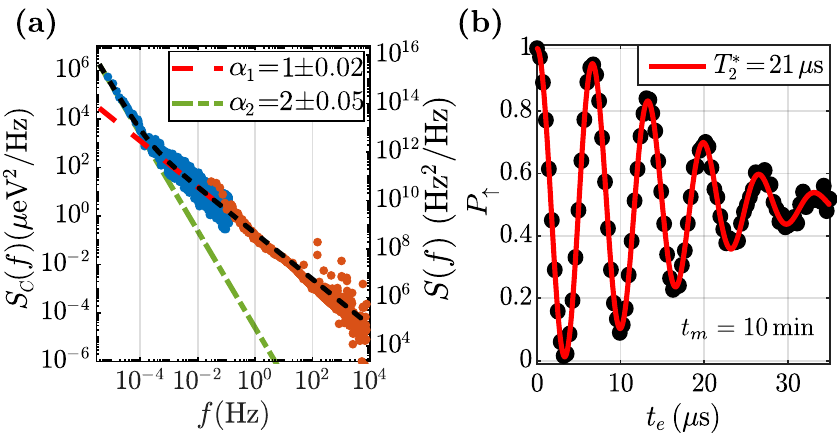}
    \caption{(a) PSD of the charge noise determined by the noise of the current $I_{\text{SET}}$ through the SET. The blue and the red dots represent two datasets. We read $S_C^{1/2}(1\text{\,Hz})=0.47$\,$\upmu$eV/$\sqrt{\text{Hz}}$ at 1\,Hz. The right y-axis is converted into the PSD scale of detuning noise by Eq. \ref{eq:conversion}. (b) Simulation of the spin-up probability as a function of the evolution time $t_e$ taking the measured charge noise spectrum into account (black dots). $T_2^*$ is fitted (red line) by the same fit function as in Fig. \ref{fig:fig2}(b).}
    \label{fig:charge_noise}
\end{figure}

In order to investigate the impact of charge noise, we measured the charge noise in the qubit vicinity via the current noise of the SET sensor. This current noise is translated into gate equivalent voltage-noise by the variation $dI_{SET}/dV_{QS}$ of the SET current by the voltage applied to the SET gate QS (see Fig. \ref{fig:measurement}(a)). We measured the current noise $\epsilon_{SET}$ at a highly sensitive operation point of the SET and subtracted from its PSD the noise spectrum measured when the SET was set to be insensitive to charge noise from the device, in order to remove noise originating from the measurement circuit\cite{Takeda13}. Fig. \ref{fig:charge_noise}(a) shows the measured PSD of the SET noise. As the data reveals two slopes, we fitted with $S_C(f)=S_{C1}/f^{\alpha_1}+S_{C2}/f^{\alpha_2}$. The fitted exponents of the $S_C(f)$ spectrum are $\alpha_1=1 \pm 0.02$ and $\alpha_2=2 \pm 0.05$, respectively, the transition being at about $10^{-3}$~Hz. This frequency dependence is in very good agreement with the one observed for the qubit detuning PSD in Fig. \ref{fig:measurement}(d) and with the qualitative trend extended to high frequencies with the Hahn-echo experiment. Making the comparison more quantitative, we assume the charge noise at the SET to be similar to the one of the QD and the longitudinal gradient magnetic field to be isotropic for lateral QD displacements. Using the current noise trace $\epsilon_{SET}(t)$, the resulting frequency detuning is

\begin{equation}
    \Delta f(t)=\epsilon_{SET}(t) \cdot \frac{dV_{QS}}{dI_{SET}}\cdot \frac{dx_{QD}}{dV_{QD}} \cdot \frac{dB_x}{dx_{QD}}\cdot \frac{g\mu_B}{\hbar}, 
    \label{eq:conversion}
\end{equation}

\noindent where $\frac{dV_{QS}}{dI_{SET}}=1/35$\,mV/pA is the inverse of the current change through the SET induced by a change of the voltage on the gate. ${dx_{QD}/dV_{QD}}=0.024$~nm/mV is the estimated displacement of the QD induced by voltage changes on the adjacent gates according to an electrostatic device simulation. ${dB_x/dx_{QD}}=0.08$~mT/nm is the simulated isotropic longitudinal gradient magnetic field at the QD position. The factor $\frac{g\mu_B}{\hbar}$, containing the electron g-factor ($g\approx2$), the Bohr magneton $\mu_B$ and the reduced Planck constant $\hbar$, converts magnetic field to frequency. We convert the charge noise PSD into qubit detuning noise by Eq. \ref{eq:conversion} (right y-axis in Fig. \ref{fig:charge_noise}(a)). The low-frequency part of the PSD (blue dots in Fig. \ref{fig:charge_noise}(a)), shows excellent agreement with the qubit detuning noise PSD $S(f)$ in its frequency dependence and its magnitude. In order to also include the high frequency range (red dots in Fig. \ref{fig:charge_noise}(a)) into the comparison, we simulated the spin-up probability after a Ramsey gate sequence at time $t_m$ with evolution time $t_e$ using 

\begin{equation}
    P_{\uparrow}(t,t_e)=\frac{1}{2}(\cos(2\pi t_e \Delta f(t))+1)
\end{equation}

\noindent and include quasi-static noise during the free evolution time $t_e$ from the full PSD in Fig. \ref{fig:charge_noise}(a). The simulated data points (black dots in Fig. \ref{fig:charge_noise}(b)) yield $T_2^*=21\,\upmu$s, which is very close to the experimentally determined value $T_2^*=18\,\upmu$s found in Fig. \ref{fig:fig2}(b). In summary, comparing data covering more than 8 frequency decades, the excellent agreement demonstrates that charge noise dominates the qubit detuning noise in our device and transits from a $S(f) \propto 1/f^2$ dependence to a $S(f) \propto 1/f$ dependence around $10^{-3}$~Hz.
 
To complete our analysis of the detuning noise and the time-ensemble spin dephasing time $T_2^*$, we estimate the magnetic noise impact due to the residual non-zero spin nuclei in our device. We can compute the resulting $T_2^*$ with (see Supplementary Note 2)

 \begin{equation}
     T_2^{*} = \frac{\hbar\sqrt{3N_S}}{p\gamma A\sqrt{2I(I+1)}},
 \end{equation}
\noindent where $N_S$ is the number of nuclei, $p$ is the fraction of nulcei with finite nuclear spin, $\gamma$ is the volume fraction of the wavefunction for which we want to calculate the infuence on $T_2^*$, i.e. localized in the barrier or the quantum well. $A$ is the hyperfine coupling constant per nucleus and $I$ is the non-zero nuclear spin.
In Ref. \cite{Hollmann_2019}, we measured the orbital splitting of this QD to be 2.5~meV. Assuming a harmonic potential, we calculate the size of the QD, taken to be the full-width-at-half-maximum of the ground state wavefunction. This yields a radius of $\approx 13$\,nm. By approximating the QD as a cylinder with height 6\,nm we estimate the number of atoms in the QD volume to be $N_A=1.6\cdot10^5$. From Schrödinger-Poisson simulations, we estimate the overlap with the SiGe barriers to be $\gamma_B\approx0.1\%$. We calculate the number of non-zero nuclear spins, which are relevant for the hyperfine coupling with the qubit (i.e. are within the the cylindrical volume assigned to the QD), for the residual 60\,ppm $^{29}$Si in the $^{28}$Si strained QW layer, residual $^{29}$Si and $^{73}$Ge in the SiGe barriers with natural abundance of isotopes as, respectively:

 \begin{align}
    N_{S,{}^{29}\mathrm{Si}}^{QW} & = p_{{}^{29}\mathrm{Si}}^\mathrm{QW}(1-\gamma_B)N_A=60\cdot10^{-6} (1-\gamma_B)N_{A} \approx 9.6,
    \\
    N_{S,{}^{29}\mathrm{Si}}^{\mathrm{barrier}} & = p_{{}^{29}\mathrm{Si}}^\mathrm{barrier}\gamma_B N_A=0.0467 \cdot 0.7 \cdot\gamma_B N_{A} \approx 5.2,
    \\
    N_{S,{}^{73}\mathrm{Ge}}^{\mathrm{barrier}} & = p_{{}^{73}\mathrm{Ge}}^\mathrm{barrier}\gamma_B N_A=0.0776 \cdot 0.3 \cdot\gamma_B N_{A} \approx 3.7.
 \end{align}

\noindent The coupling constants are $A_{\text{Si}}=2.15\,\upmu$eV and  $A_{\text{Ge}}\approx 10\cdot A_{\text{Si}}$ \cite{PhysRev.134.A265, PhysRevB.85.205312}, respectively, with $I_{\text{$^{29}$Si}}=1/2$, $I_{\text{$^{73}$Ge}}=9/2$. Assuming the spin baths to be in the ergodic limit, each subset of nuclear spin results in the following dephasing times:
$T_2^*(_{^{29}Si}^{\text{QW}})  = 22\,\upmu\text{s}$, $T_2^*(_{^{29}Si}^{\text{barrier}}) = 30\,\upmu\text{s}$ and $T_2^*(_{^{73}Ge}^{\text{barrier}}) = 0.61\,\upmu\text{s}$.

\noindent Notably, due to the strong hyperfine coupling of the $^{73}$Ge in the barrier layers, the $^{73}$Ge alone would dephase the qubit faster than observed in the experiment shown in Fig. \ref{fig:fig2}(c). This apparent contradiction is resolved, if the correlation time of the $^{73}$Ge nuclear spin bath is larger than a few hours and thus the ergodic limit is not reached in our $T_2^*(t_m)$ measurement. 

In conclusion, we have shown that in a highly purified $^{28}$Si/SiGe qubit device with 60\,ppm residual $^{29}$Si, the $^{73}$Ge nuclear spins in the potential barrier do not dominantly contribute to the qubit dephasing time, despite their strong hyperfine coupling. We find the dynamics of the nuclear spin bath to be slower than 6~hours, similarly to observations for electrons bound to single phosphorus donors in 800\,ppm residual $^{29}$Si in the presence of the electron's Knight shift \cite{Mateusz_2019}. Thus, the improvement potential of qubit dephasing times that can be expected from isotopical purification of the natural SiGe barrier is negligibly weak. In our device featuring a nanomagnet integrated into the gate layout for EDSR manipulation, charge noise is the dominant qubit noise source in a frequency range of more than 8 decades. In the low frequency regime, the charge and the qubit detuning noise present a $1/f^2$ dependence below $1\cdot10^{-3}$\,Hz. Above, towards higher frequencies, both PSD transit to a $1/f$ dependence. This $1/f$ trend was recently also observed in a device featuring a micromagnet and 800\,ppm $^{29}$Si \cite{tarucha_quantum-dot_2017}. From the Hahn-echo experiment for our qubit detuning, we additionally deduce a remarkably high $T_2^{\text{echo}}=128\,\upmu$s. We finally show $T_2^*$ to clearly and monotonously decrease for measurement times increasing from seconds to several hours, in accordance with the absence of a roll-off in the charge noise PSD down to at least $5\cdot10^{-5}$\,Hz. Our experimental $T_2^* \approx 18\,\upmu$s for $t_m=600$~s quantitatively results from the charge noise $S_C^{1/2}(1\text{\,Hz})=0.47$\,$\upmu$eV/$\sqrt{\text{Hz}}$, which falls within the range of 0.3 to 2 $\upmu$eV/$\sqrt{\text{Hz}}$ seen in literature \cite{Freeman_2016,Connors_2019,Petit_2018,Mi_2018}. While the on-chip integration of a micro- or nanomagnet does not induce additional magnetic noise \cite{Neumann15, tarucha_quantum-dot_2017}, minimizing the newly opened electric dephasing channel seems to be key for further significant improvement of spin qubit gate fidelities in highly purified $^{28}$Si compared to devices avoiding integrated static magnetic field gradients\cite{Thorgrimsson17, Huang2019, Yang19, Andrews19}.

See supplementary material for details of the measurement cycle, simulations of the nanomagnet and derivations for the $T_2^*$ time in the ergodic limit and its dependence on measurement time. 

This work has been funded by the German Research Foundation (DFG) within the projects BO 3140/4-1, 289786932 and the cluster of excellence "Matter and light for quantum computing" (ML4Q) as well as by the Federal Ministry of Education and Research under Contract No. FKZ: 13N14778. Project Si-QuBus received funding from the QuantERA ERA-NET Cofund in Quantum Technologies implemented within the European Union's Horizon 2020 Programme. 

\bibliography{DetuningNoiseBiblio_clean}

\clearpage



\end{document}